# Polymorphic crystallites model for monolayer amorphous materials

Le-Ye Zhu (朱乐烨)[1,2#], Xi Zhang (张熙) [1,2#], Yun-Peng Wang (王云鹏)[3], Jieheng Shi (施杰亨)[1], Junwei Zhang (张俊炜)[1], Shixuan Du (杜世萱)[1,2,4], and Yu-Yang Zhang (张余洋)[1*]

1 School of Physical Sciences, University of Chinese Academy of Sciences, Beijing 100049, China

2 Institute of Physics, Chinese Academy of Sciences, Beijing 100190, China

3 School of Physics, Central South University, Changsha 410083, China

4 Songshan Lake Materials Laboratory, Dongguan, Guangdong 523808, China

* Authors to whom correspondence should be addressed: zhangyuyang@ucas.ac.cn

## ABSTRACT

Establishing atomistic models of amorphous materials remains a central challenge in materials science. Here, we examine three representative multicomponent monolayer systems using first-principles-trained machine-learning potentials (MLPs) and an energy-guided Monte Carlo structural-search workflow. For monolayer amorphous boron nitride (maBN), independent density-functional-theory (DFT) single-point calculations confirm that the ensemble generated by the MLP-driven search is lower in energy on average than an extended-Tersoff-generated ensemble. Across the sampled maBN structures, local-ring classification combined with strict connected-domain analysis identifies h-BN-like, o-$B_2N_2$-like, and o-$B_4N_4$-like crystallite domains together with mixed medium-range order. Only individual samples containing at least two distinct phase-pure crystallite-domain types are classified as polymorphic crystallites. The same analysis finds coexisting hexagonal and tetragonal domains in monolayer amorphous LiCl and graphene-like, h-BN-like, borophene-like, and mixed-order regions in monolayer amorphous BCN. Five-nanosecond MLP molecular-dynamics trajectories at 300 K show that the candidate structures are dynamically persistent on the simulated timescale. These results support a polymorphic-crystallite description for the selected model systems, while experimental discrimination from alternative continuous-random-network descriptions will require medium-range-order-sensitive measurements.

Amorphous materials are widespread, structurally diverse forms of matter whose disordered configurations enable a broad range of applications[1–8]. Their lack of long-range order, however, makes atomic structure difficult to determine using diffraction alone. An atomistic description of amorphous materials therefore remains a major challenge. Two historically important descriptions are the continuous random network (CRN) proposed by Zachariasen in 1932[9] and the earlier crystallite model of Lebedev[10]. The debate between these descriptions persisted through the mid-twentieth century[11], and has been renewed by advances in characterization and simulation[12–20]. Different preparation conditions can produce distinct structural forms of amorphous silicon[21], while amorphous diamond has been reported with either crystallite-containing or predominantly CRN-like structures[22,23]. These findings show that amorphous topology can depend on preparation history. They also emphasize that agreement with a radial distribution function, although necessary, cannot by itself uniquely resolve competing medium-range-order models.

In 2020, Toh et al. reported monolayer amorphous carbon (MAC), enabling direct atomic-scale structural analysis by high-resolution transmission electron microscopy (HRTEM)[24]. Subsequent work showed that the extent of ordered regions in MAC can be altered by preparation conditions and correlated with electrical and mechanical responses[25,26]. Multicomponent two-dimensional amorphous materials introduce chemical decoration in addition to topology, and therefore offer a wider motif space.

Previous simulations of monolayer amorphous boron nitride (maBN) described noncanonical six-membered rings containing B–B and N–N bonds as pseudocrystallite regions[27,28], while experiment has found a random nitrogen distribution in nitrogen-doped amorphous monolayer carbon[29]. A reproducible distinction among local motifs, connected crystallite domains, mixed medium-range order, and a CRN is therefore needed before a structural model can be assigned to a multicomponent monolayer.

In this work, we use an EDKMC-based active-learning workflow to generate first-principles training data and to search candidate structures for maBN, monolayer amorphous LiCl (maLiCl), and monolayer amorphous BCN (maBCN)[30]. Here, a CRN contains no spatially continuous domain assignable to a crystalline reference tiling, although local coordination order may remain. A local motif is a chemically and topologically specified polygonal unit compatible with such a reference tiling. A crystallite denotes a connected domain whose local bonding, chemical decoration, and tiling relation match one crystalline reference. A pseudocrystallite-like region possesses local or medium-range order but does not satisfy the complete reference-template, connectivity, and minimum-domain criteria. A polymorphic-crystallite structure contains two or more crystallographically distinct crystallite-domain types within the same amorphous network. Within these definitions, the calculations predict DFT-consistent candidate polymorphic-crystallite and mixed-order structures that persist dynamically over the simulated timescale. The three systems test the applicability of this classification to representative multicomponent monolayers; they do not establish a universal or experimentally unique amorphous structure.

The trained MLPs were used both in molecular dynamics (MD) and in energy-driven kinetic Monte Carlo (EDKMC) structural searches. In the present application, EDKMC is not interpreted as a physical annealing or synthesis trajectory. Stone–Wales and atom-exchange moves propose topological or chemical rearrangements, and each trial is accepted from its energy change $\Delta E$ according to $P_{acc} = 1$ for $\Delta E \leq 0$ and $P_{acc} = \exp(-\Delta E/\chi)$ for $\Delta E > 0$. The quantity $\chi$ is an algorithmic acceptance-energy parameter that controls uphill acceptance; it has no thermodynamic-temperature meaning and is not mapped to a synthesis variable. Using EDKMC within the active-learning loop expands sampling beyond the time-accessible pathways of conventional MD while retaining DFT labeling of configurations selected by model disagreement.

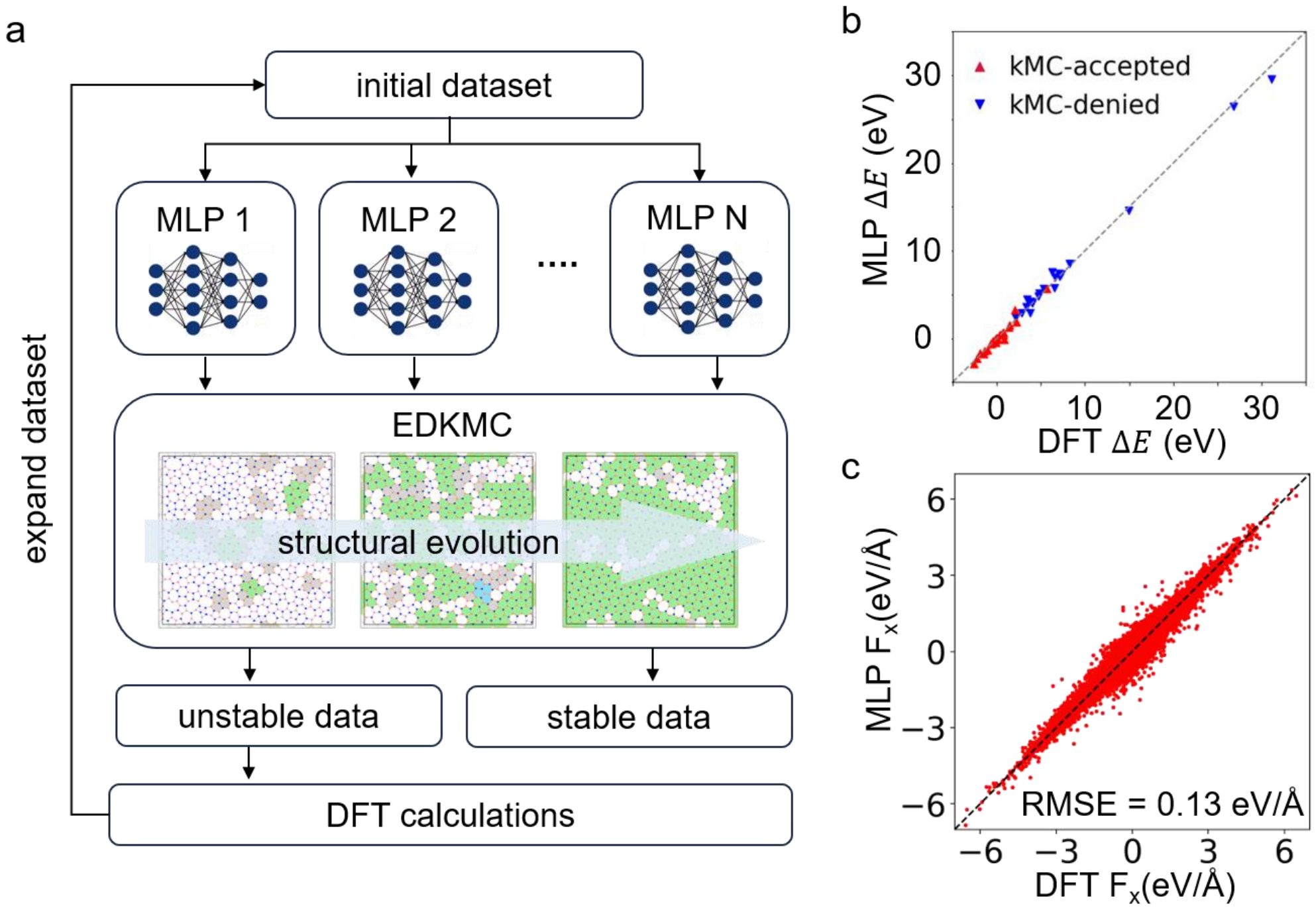

**Figure 1. EDKMC-based active-learning workflow, illustrated for maBN.** (a) Schematic of the iterative workflow. (b) Comparison of energy changes caused by Stone–Wales transformations and atom exchanges for 50 accepted steps (red triangles) and 50 rejected steps (blue inverted triangles), calculated by DFT and the MLP. (c) Comparison of atomic forces in held-out amorphous configurations calculated by DFT and the MLP.

Figure 1(a) summarizes the active-learning procedure. An initial dataset containing crystalline structures and defective configurations, including Stone–Wales defects, atom exchanges, and vacancies, is labeled by DFT. Four MLPs with different initial parameters are trained and then used to drive EDKMC structural searches. Disagreement among their energy predictions identifies configurations that are poorly represented in the current training set; these configurations are recalculated by DFT and added to the dataset. The cycle is repeated until the model-deviation criterion no longer selects new configurations (see the Methods in the Supporting Information). The structures shown in Fig. 1(a) are search-generated candidates with different connected structural-order fractions, not snapshots of a physical thermal-annealing process.

This procedure produced MLPs for maBN, maLiCl, and maBCN. The held-out per-atom energy root-mean-square errors are 8.3, 8.5, and 12 meV atom$^{-1}$, respectively, and the corresponding force errors are 0.13, 0.046, and 0.21 eV Å$^{-1}$. Figure 1(b) further tests the quantity that directly controls an EDKMC step: the energy difference between the trial and current configurations. For the sampled accepted and rejected maBN moves, MLP and DFT energy differences agree closely. Independent 15-step comparisons for all three systems, together with the held-out energy and force parity plots, are reported in Sec. II of the Supporting Information.

Validation was extended beyond scalar errors. Independent NVT trajectories at 300 K were run with DFT and the corresponding MLP from the same 60-atom structures and equivalent initial velocities. The partial in-plane radial distribution functions (RDFs), bond-length distributions, and bond-angle distributions agree closely for all three systems. For a larger maBN model, the MLP-MD first-neighbor distance is approximately 1.44 Å, qualitatively consistent with reported finite-thickness amorphous BN films[31]. RDFs and structure factors provide two-body structural descriptors, but do not uniquely distinguish a CRN, mixed medium-range order, and connected crystallite domains; this limitation is retained in interpreting the structures below.

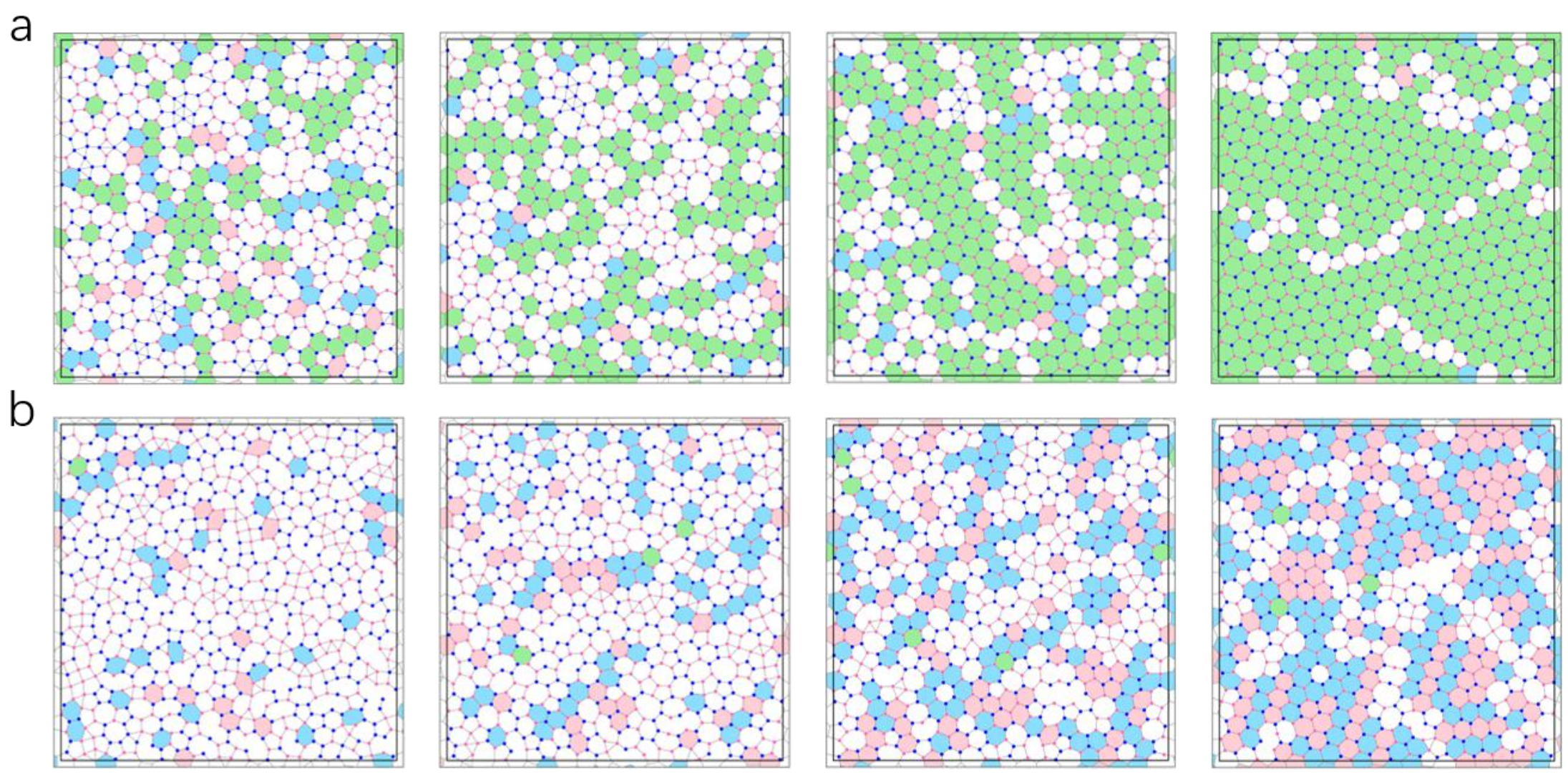

**Figure 2. Representative metastable maBN configurations.** (a) maBN candidates generated by the MLP-driven search and (b) maBN candidates generated by the extended-Tersoff-driven search. Local h-BN-like, o-$B_2N_2$-like, and o-$B_4N_4$-like six-membered rings are colored green, pink, and blue, respectively. The colors identify local ring classes and do not by themselves assign connected crystallite domains.

With the validated MLP, we performed EDKMC structural searches for maBN and compared them with searches driven by the extended Tersoff potential at the same acceptance parameter[32]. Figure 2(a) shows representative MLP-generated configurations rich in h-BN-like rings, whereas Fig. 2(b) shows extended-Tersoff-generated configurations with many noncanonical rings containing B–B and N–N bonds. These configurations are metastable search outcomes, not a simulation of thermal annealing or proof of a unique equilibrium amorphous structure. To avoid a circular energy comparison, 15 equilibrated configurations from each search were recalculated by identical DFT single-point calculations without further relaxation. The sampled MLP-generated ensemble has a DFT mean energy of $-8.49 \pm 0.06$ eV atom$^{-1}$, compared with $-7.60 \pm 0.13$ eV atom$^{-1}$ for the sampled extended-Tersoff-generated ensemble. The independent DFT benchmark therefore confirms the lower mean energy of the sampled MLP-generated ensemble relative to the sampled extended-Tersoff-generated ensemble (Fig. S8). This comparison concerns the two search-generated ensembles and does not establish an intrinsic energetic ordering among polymorphic-crystallite, pseudocrystallite-like, and CRN structural classes.

**Table 1. Local six-membered-ring census for the four extended-Tersoff-generated maBN samples shown in Fig. 2(b).** A and B denote the two atomic species; sequences related by cyclic permutation, reversal, or A/B exchange belong to the same equivalence class. The counts are local ring statistics and are not crystallite-domain area fractions.

| **Six-membered-ring class** | Sample 1 | Sample 2 | Sample 3 | Sample 4 |
|---|---|---|---|---|
| AABABB(o-$B_4N_4$) | 32 | 54 | 97 | 135 |
| AABAAB(o-$B_2N_2$) | 18 | 33 | 63 | 112 |
| ABABAB(h-BN) | 1 | 3 | 5 | 3 |
| Other | 48 | 57 | 37 | 21 |
| Total hexagons | 99 | 147 | 202 | 271 |

The ring census separates local chemical decoration from domain assignment. Under the stated equivalence operations, ABBABB and AABAAB belong to the o-$B_2N_2$-like class, ABBAAB and AABABB belong to the o-$B_4N_4$-like class, and ABABAB is h-BN-like; all remaining decorations are grouped as "other." Table 1 shows that the extended-Tersoff-generated samples are not described by a purely random B/N decoration at the local-ring level. Nevertheless, an isolated classified ring is only a local motif and is not, by itself, a crystallite.

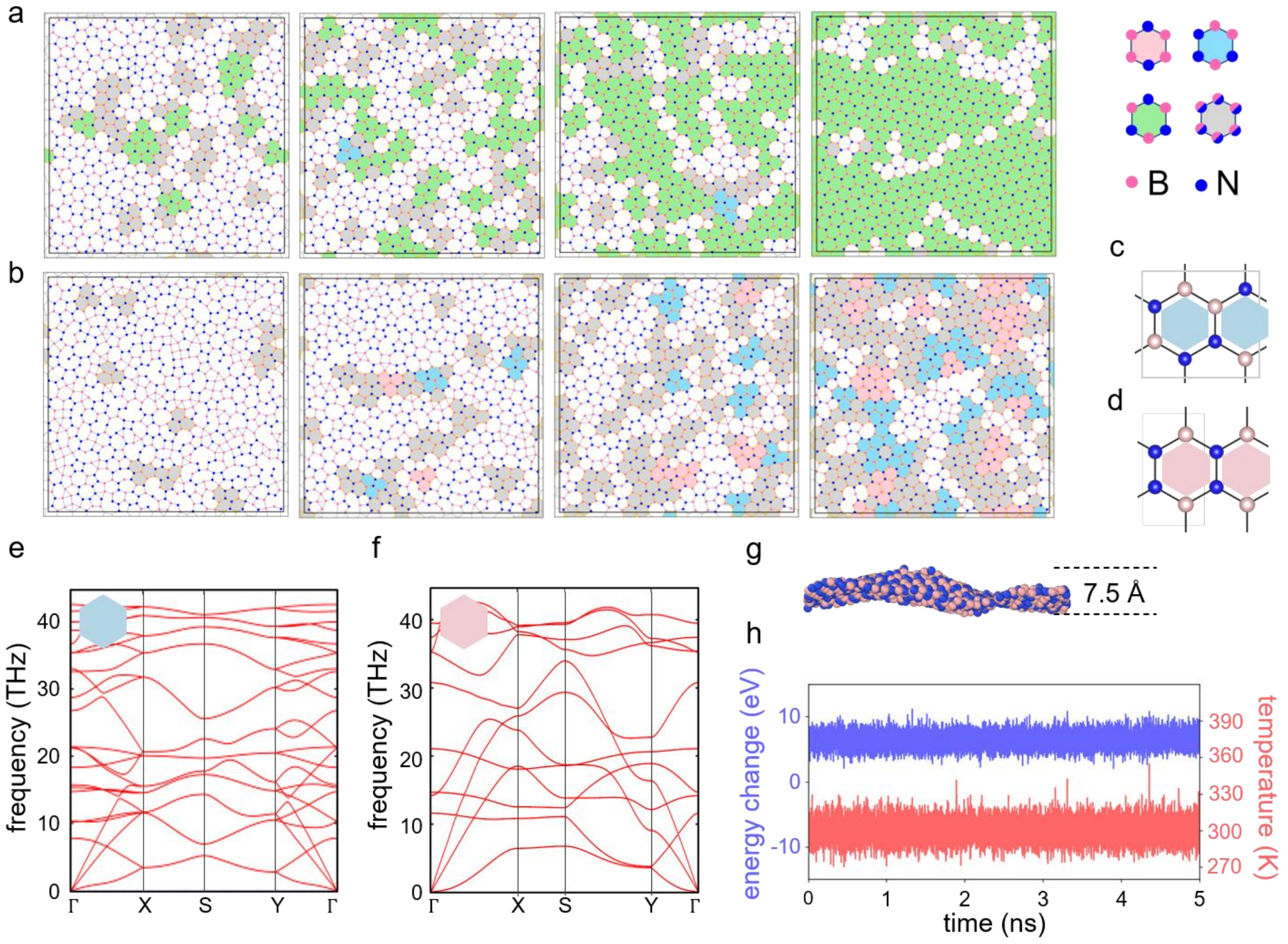

**Figure 3. Connected-domain analysis and dynamical persistence of maBN.** (a,b) Domain maps of the structures in Fig. 2 using the strict connected-domain rules; h-BN-like, o-$B_2N_2$-like, o-$B_4N_4$-like, and other hexagons are colored green, pink, blue, and gray. (c,d) Unit cells of periodic o-$B_4N_4$ and o-$B_2N_2$. (e,f) Their phonon dispersions, which test only the ideal periodic reference phases. (g) Side view of the extended-Tersoff-generated maBN structure after 5 ns of MLP-MD at 300 K. (h) Energy and temperature during that trajectory.

Figures 3(a) and 3(b) apply a second, connectivity-based stage. A minimum honeycomb domain contains three compatible hexagons in a compact pairwise edge-sharing patch; isolated hexagons and two-hexagon clusters are excluded. Larger domains grow by edge sharing while preserving the reference tiling. A component made from one ring class is assigned as a phase-pure crystallite domain, whereas a connected component that contains compatible motifs from more than one class is labeled mixed medium-range order (mixed MRO), not a phase-pure crystallite.

**Table 2. Phase-resolved connected-domain areas, $F_{MRO}$, and $DOD_{str}$ for the MLP-generated and extended-Tersoff-generated maBN samples in Fig. 3.**

| Domain type or metric | MLP-1 | MLP-2 | MLP-3 | MLP-4 | Ter-1 | Ter-2 | Ter-3 | Ter-4 |
|---|---|---|---|---|---|---|---|---|
| o-$B_4N_4$ | 0 | 3 | 4 | 0 | 0 | 11 | 26 | 65 |
| o-$B_2N_2$ | 0 | 0 | 0 | 0 | 0 | 6 | 16 | 55 |
| h-BN | 33 | 87 | 184 | 290 | 0 | 0 | 0 | 0 |
| Mixed | 63 | 70 | 43 | 7 | 30 | 63 | 131 | 144 |
| $F_{MRO}$ | 24.9% | 42.0% | 60.9% | 78.9% | 7.8% | 21.0% | 45.5% | 69.4% |
| $DOD_{str}$ | 75.1% | 58.0% | 39.1% | 21.1% | 92.2% | 79.0% | 54.5% | 30.6% |

According to the definition above, MLP-2 and MLP-3 contain both h-BN-like and o-$B_4N_4$-like crystallite domains and therefore satisfy the polymorphic-crystallite criterion. Although their o-$B_4N_4$-like areas are small, the corresponding components satisfy the connectivity and minimum-domain criteria applied throughout the analysis. Ter-2, Ter-3, and Ter-4 contain both o-$B_2N_2$-like and o-$B_4N_4$-like crystallite domains and likewise satisfy this criterion. In contrast, MLP-1 and MLP-4 contain only h-BN-like phase-pure domains together with mixed MRO, whereas Ter-1 contains only mixed MRO; these samples are therefore not classified as polymorphic crystallites. No individual sample in Table 2 contains all three phase-pure maBN domain types simultaneously.

The connected structural-order fraction is defined as $F_{MRO} = A_{MRO}/A_{tot}$, where $A_{MRO}$ is the union of the projected polygon areas belonging to accepted phase-pure and mixed-MRO components and $A_{tot}$ is the projected cell area. The structural degree of disorder is $DOD_{str} = 1 - F_{MRO}$. For maBN, $F_{MRO}$ is partitioned into h-BN-like, o-$B_2N_2$-like, o-$B_4N_4$-like, and mixed-MRO contributions, which are reported separately in Table 2. Thus, $F_{MRO}$ measures connected structural order, whereas the phase-resolved areas specify its crystallographic character. $DOD_{str}$ is used only as a geometrical descriptor and is not a thermodynamic or mechanical stability criterion.

The o-$B_2N_2$[33] and o-$B_4N_4$ motifs can tile ideal periodic honeycomb monolayers, whose unit cells and phonon dispersions are shown in Figs. 3(c)–3(f). The absence of imaginary phonon modes establishes the dynamical stability of these ideal periodic reference phases only. Relative to h-BN, DFT places o-$B_2N_2$ and o-$B_4N_4$ at 1.65 and 0.81 eV atom$^{-1}$, respectively; the MLP gives 1.63 and 0.84 eV atom$^{-1}$, while the extended Tersoff potential reverses the DFT ordering and gives −0.51 and −0.27 eV atom$^{-1}$. The MLP therefore reproduces the DFT energetic hierarchy, whereas the empirical potential over-stabilizes noncanonical environments containing B–B and N–N bonds. This difference explains the much larger population of such local motifs in the extended-Tersoff-generated structures.

We further tested the dynamical persistence of the extended-Tersoff-generated maBN structure by a 5 ns NVT MLP-MD trajectory at 300 K. No irreversible reconstruction or collapse was observed within the simulation interval, and the energy and temperature fluctuate around stationary values [Figs. 3(g) and 3(h)]. This result shows kinetic persistence in the sampled metastable basin on the simulated timescale; it does not demonstrate thermodynamic stability. Likewise, the periodic-phase phonons do not constitute a vibrational-stability test of the entire amorphous network. Experimental discrimination of the predicted medium-range domains from alternative CRN-like descriptions will require a medium-range-order-sensitive measurement such as fluctuation electron microscopy or atom-resolved HRTEM.

To examine the applicability of the same framework beyond maBN, we applied the structural-search and domain-classification protocol to monolayer LiCl. In the search shown in Fig. 4(a), Stone–Wales transformations and atom exchanges drive the system from the crystalline starting structure toward a stationary energy range after approximately 5000 accepted steps. A separate random-start control approaches a similar stationary energy range (Fig. S9), reducing the possibility that the coexistence described below is inherited solely from the crystalline initial topology.

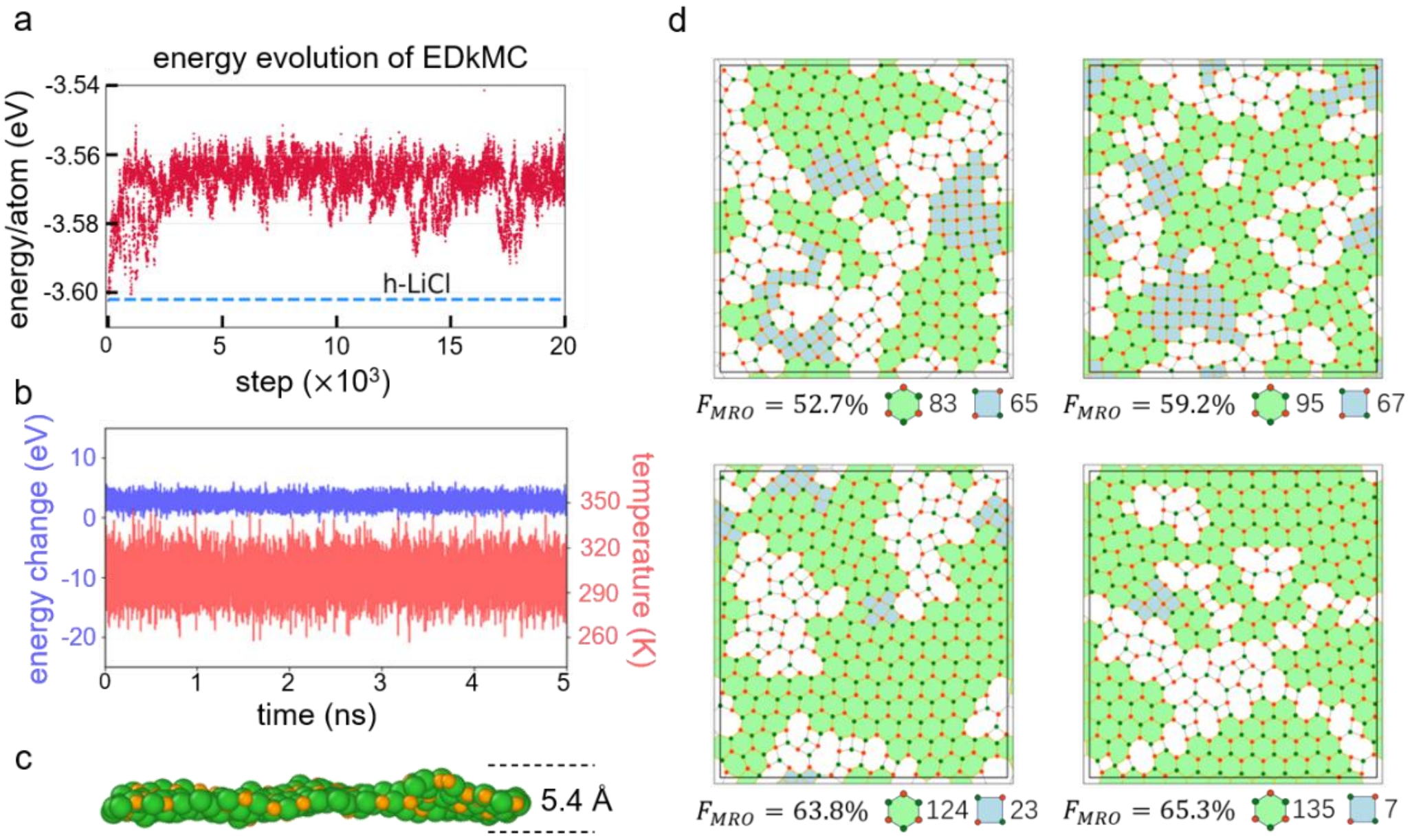

**Figure 4. Structural search and dynamical persistence of monolayer amorphous LiCl.** (a) Energy evolution during the EDKMC search; the energy of crystalline h-LiCl is shown by the blue dashed line. (b) Energy and temperature during MLP-MD at 300 K. (c) Side view after 5 ns of MLP-MD. (d) Representative maLiCl configurations at different search stages. Blue and green regions denote connected hexagonal and tetragonal domains, respectively, with the corresponding area fractions labeled below each structure.

The four maLiCl configurations in Fig. 4(d) contain a CRN-like background together with randomly oriented connected domains. Two phase-pure motif classes are resolved: hexagonal LiCl domains (blue) and tetragonal LiCl domains (green). The ideal periodic forms of both motifs have been predicted to be dynamically stable[34-36], providing crystalline templates for the domain analysis rather than a stability proof for the amorphous network. The connected ordered area generally increases along the displayed structural-search sequence. Because the main search starts from h-LiCl and conserves density, the quantitative hexagonal-to-tetragonal ratio remains conditional on the fixed density, move set, acceptance parameter, and finite search history, even though the random-start control reaches a similar stationary energy range.

The maLiCl configuration with $F_{MRO} = 52.7\%$ was propagated for 5 ns by NVT MLP-MD at 300 K. Its energy remains stationary and no irreversible structural collapse is observed [Figs. 4(b) and 4(c)]. The final structure has a maximum thickness of approximately 5.4 Å. These observations establish dynamical persistence under the simulated conditions, not thermodynamic stability of maLiCl or of the polymorphic-crystallite description.

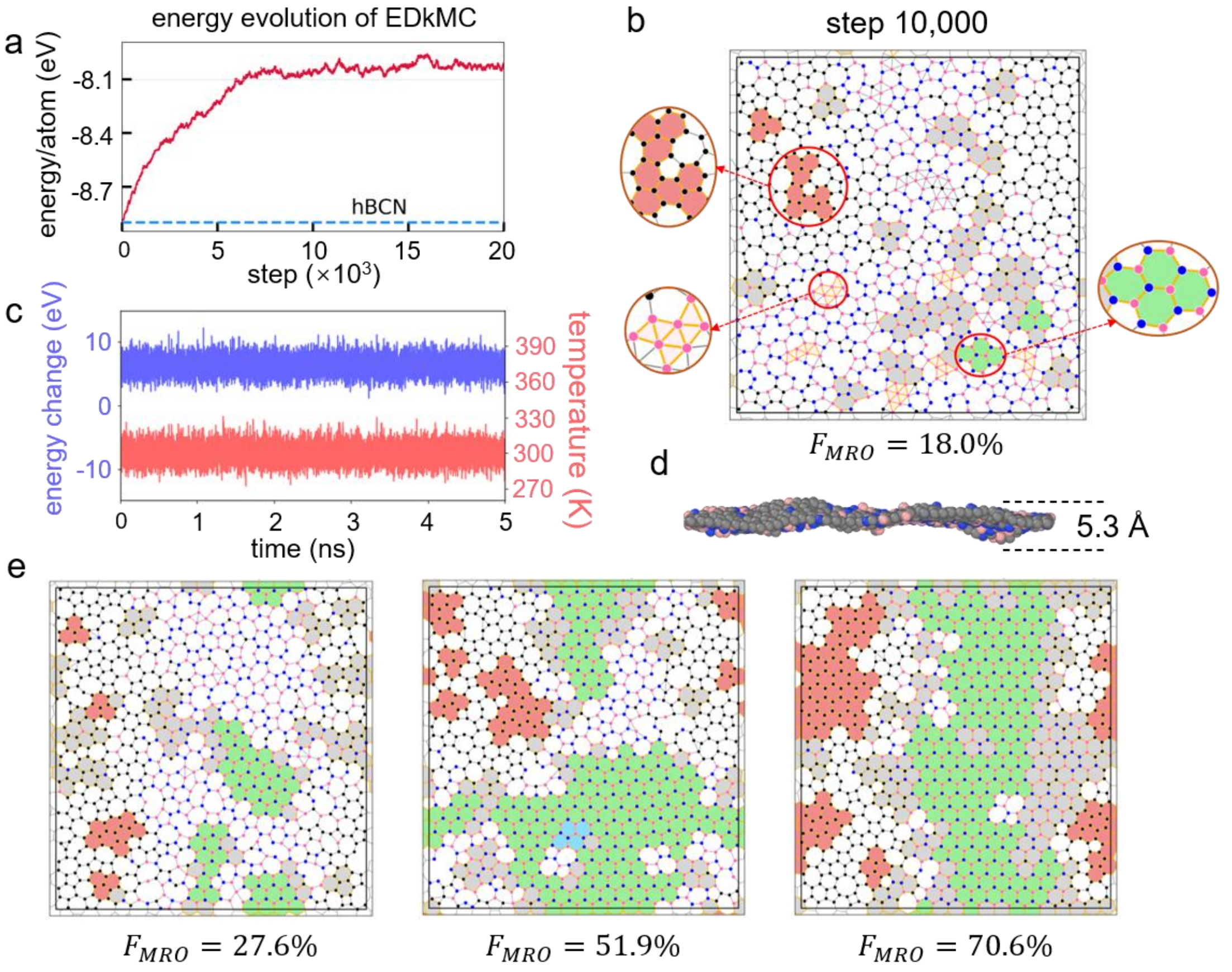

**Figure 5. Structural search and dynamical persistence of monolayer amorphous BCN.** (a) Energy evolution during the EDKMC search; the crystalline h-BCN starting energy is shown by the blue dashed line. (b) Representative maBCN configuration, with insets identifying different local motif classes. (c) Energy and temperature during MLP-MD at 300 K. (d) Side view after 5 ns of MLP-MD. (e) Candidate configurations spanning different $DOD_{str}$ values; identified connected ordered regions are shown in green, with their area fractions labeled below.

We next applied the framework to ternary maBCN. The search in Fig. 5(a) starts from an h-BCN configuration in which BN and C are segregated into stripes and approaches a stationary energy range after approximately 10 000 steps. To test sensitivity to this chemical seeding, additional searches were initialized from random, well-mixed, and island-like arrangements (Fig. S10). We monitored the chemical-order parameter P = ($N_{C–C}$ + $N_{B–N}$)/$N_{total}$, where a larger P denotes greater C–C/B–N ordering. Under the stated acceptance parameter and move set, the well-mixed and island-like searches approach comparable stationary energies and P values. These controls reduce a simple seeding explanation, but the detailed pathways, motif fractions, and domain morphology remain conditional on the initial topology and finite structural-search protocol.

The representative candidate in Fig. 5(b) contains graphene-like domains formed by pure-carbon hexagons, h-BN-like domains, triangular borophene-like motifs, and connected mixed-MRO regions with B/C/N decoration. The coexistence of several crystalline-reference motif classes is consistent with the polymorphic-crystallite classification, but it does not establish that maBCN spontaneously selects a unique equilibrium domain pattern. Rather, the calculations identify DFT-consistent metastable candidates accessible to the adopted structural-search protocol.

A 5 ns NVT MLP-MD trajectory at 300 K was used to test the dynamical persistence of the maBCN candidate. The energy and temperature remain stationary [Fig. 5(c)], while the final side view [Fig. 5(d)] shows no irreversible reconstruction and only limited out-of-plane corrugation within the simulation interval. As for maBN and maLiCl, this result demonstrates kinetic persistence on the simulated timescale and should not be interpreted as thermodynamic stability.

Varying the algorithmic acceptance parameter $\chi$ generates a controlled ensemble of maBCN candidate configurations with different connected structural-order fractions [Fig. 5(e)]. Their structural disorder is reported as $DOD_{str} = 1 - F_{MRO}$ using the same connected-component and projected-area procedure as for maBN and maLiCl. Because $\chi$ has no thermodynamic-temperature interpretation, these calculations do not demonstrate experimental tunability or define a mapping between $DOD_{str}$ and a synthesis condition; they show only that the structural-search protocol samples candidates spanning a range of motif fractions and connected order.

In summary, an EDKMC-based active-learning workflow was used to construct MLPs and to perform energy-guided structural searches for three representative multicomponent monolayer systems. For maBN, independent DFT single-point calculations confirm that the sampled MLP-generated candidate ensemble is lower in energy on average than the sampled extended-Tersoff-generated ensemble; this ensemble comparison is not interpreted as an intrinsic stability ordering among structural classes. Separating local ring classification from strict connected-domain assignment resolves h-BN-like, o-$B_2N_2$-like, o-$B_4N_4$-like, and mixed-MRO components and explains why local ring counts cannot be equated with crystallite area. The same classification identifies coexisting hexagonal and tetragonal domains in maLiCl and graphene-like, h-BN-like, borophene-like, and mixed-order regions in maBCN. The 5 ns MLP-MD trajectories demonstrate dynamical persistence of these metastable candidates under the simulated conditions. The sampled structures include both polymorphic-crystallite candidates and non-polymorphic mixed-order candidates. The identified polymorphic examples support the applicability of a DFT-consistent polymorphic-crystallite description to selected structures, rather than a universal label for every member of either search-generated ensemble or an experimentally established paradigm. Distinguishing these predicted medium-range domains from alternative CRN-like descriptions remains a task for fluctuation electron microscopy or atom-resolved HRTEM.

## Supporting Information

See the Supporting Information for details of the DFT calculations, MLP training, and EDKMC structural-search protocol; held-out and finite-temperature MLP validation; local-motif and connected-domain definitions; independent DFT energy benchmarks for maBN; initial-condition and acceptance-parameter controls.

## Acknowledgment

This work was supported by the National Natural Science Foundation of China (No. 92477205, No. 22372047, and No. 52250402), the National Key R&D Program of China (No. 2024YFA1207800), CAS Project for Young Scientists in Basic Research (YSBR-003), and Beijing Natural Science Foundation (QY25168).

## AUTHOR DECLARATIONS

### Conflict of Interest

The authors have no conflicts to disclose.

### Author Contributions

[#]Le-Ye Zhu and Xi Zhang contributed equally.

## DATA AVAILABILITY

The data that support the findings of this study are available from the corresponding authors upon reasonable request.

## REFERENCE：

# Supporting Information for

# "Polymorphic crystallites model for monolayer amorphous materials"

Le-Ye Zhu (朱乐烨)[1,2#], Xi Zhang (张熙)[1,2#], Yun-Peng Wang (王云鹏)[3], Jieheng Shi (施杰亨)[1], Junwei Zhang (张俊炜)[1], Shixuan Du (杜世萱)[1,2,4], and Yu-Yang Zhang (张余洋)[1*]

[1] School of Physical Sciences, University of Chinese Academy of Sciences, Beijing 100049, China

[2] Institute of Physics, Chinese Academy of Sciences, Beijing 100190, China

[3] School of Physics, Central South University, Changsha 410083, China

[4] Songshan Lake Materials Laboratory, Dongguan, Guangdong 523808, China

* Author to whom correspondence should be addressed: zhangyuyang@ucas.ac.cn

## I. Methods

### A. Density-functional-theory calculations

Density-functional-theory (DFT) calculations were performed with the Vienna Ab initio Simulation Package (VASP)[1]. Exchange and correlation were treated using the Perdew–Burke–Ernzerhof generalized-gradient approximation, and the projector-augmented-wave method was used[2]. Plane-wave cutoff energies of 500, 600, and 700 eV were adopted for BN, LiCl, and BCN, respectively. Total energies were evaluated using Γ-centered k-point meshes generated with a reciprocal-space density of 6 $Å^{-1}$. The same element-specific settings were used for the training configurations, MC-step benchmarks, and independent single-point tests.

### B. Machine-learning potentials

The training sets were generated iteratively with the DP-GEN active-learning workflow[3]. Energy-guided Monte Carlo structural searches were used to sample configurations spanning different topologies and structural disorder, after which configurations identified by the model-deviation criterion were labeled by DFT and added to the training pool. The final machine-learning potentials (MLPs) used the smooth Deep Potential model DeepPot-SE (se_e2_a)[4], as implemented in DeePMD-kit[5,6].

The neighbor cutoff was 6.5 Å, with the smoothing function beginning at 4.0 Å. The embedding network contained hidden layers of 25, 50, and 100 neurons, and the fitting network contained three hidden layers of 120 neurons each. For each chemical system, 90% of the labeled configurations were selected randomly for training and the remaining 10% were held out for validation.

The final maBN, maLiCl, and maBCN models were trained on 105,870, 31,624, and 121,889 configurations, respectively. The corresponding periodic cells were 13 × 13 × 20 $Å^3$ for maBN, 20

× 20 × 20 Å³ for maLiCl, and 10 × 10 × 20 Å³ for maBCN.

### C. Energy-driven Monte Carlo structural search

The energy-driven kinetic Monte Carlo (EDKMC) procedure is used here as a nonphysical, energy-guided structural-search protocol rather than as a model of synthesis, annealing, or an amorphization pathway[7]. This choice avoids the irreversible out-of-plane collapse observed when the monolayers are melted and quenched. The search samples low-lying metastable basins that are accessible through the selected topological moves; its trajectory is not assigned a physical time.

Each trial consists of either a Stone–Wales bond rotation or an atom-exchange move, followed by local relaxation with the relevant MLP. Writing $\Delta E = E_{new} - E_{old}$, a relaxed trial is accepted with $P_{acc} = 1$ for $\Delta E \leq 0$ and $P_{acc} = \exp\left(-\frac{\Delta E}{\chi}\right)$ for $\Delta E > 0$. Here $\chi$ is an algorithmic acceptance-energy parameter. Values from 0.5 to 3.0 eV were used to generate candidate ensembles with different structural characteristics. $\chi$ has no thermodynamic-temperature interpretation and is not mapped to a synthesis variable. The symbol $k_B T$ that remains in the legends of Fig. S10 denotes this energy-like acceptance parameter only.

## II. Validation of the machine-learning potentials

### A. Held-out energy and force errors

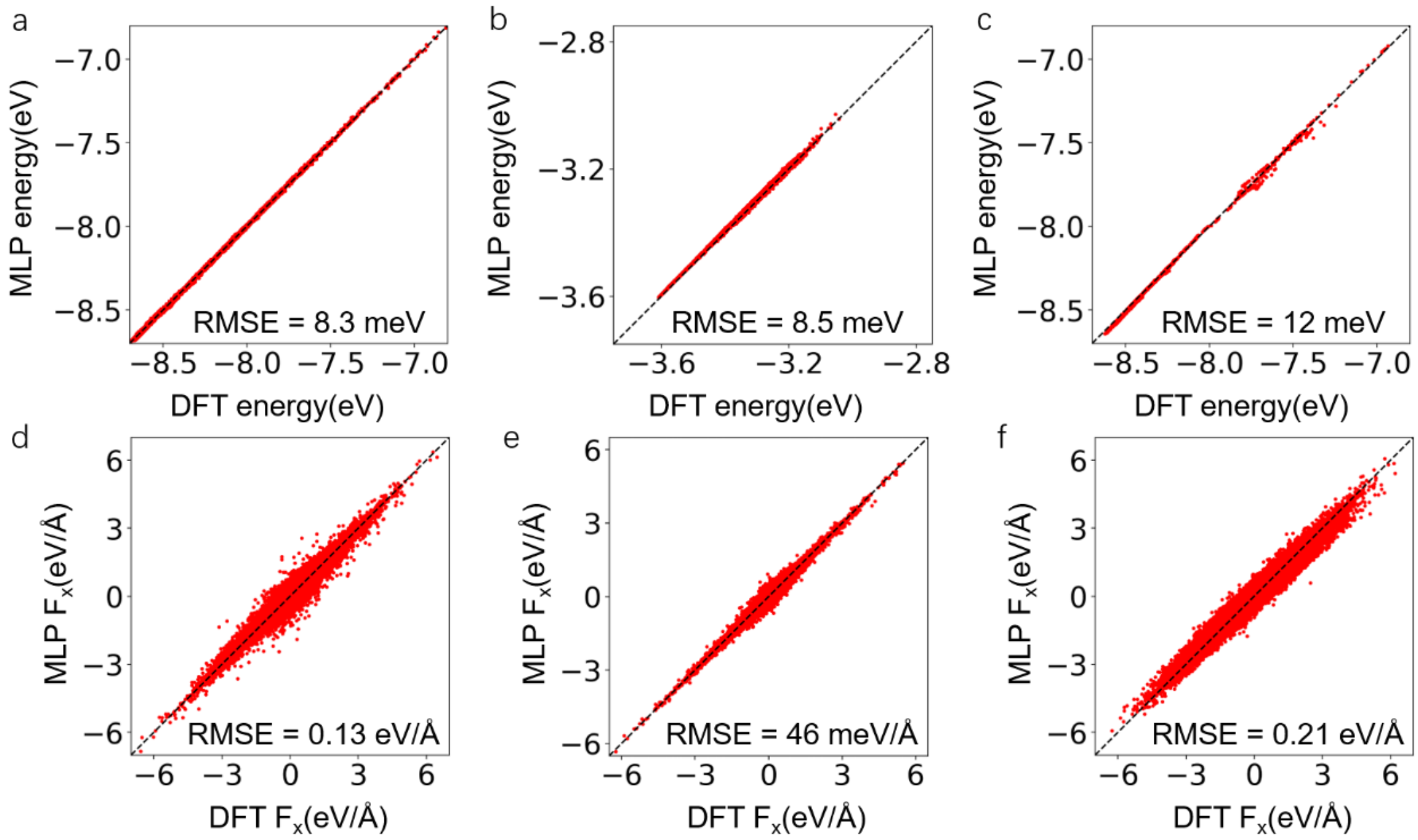

**Figure S1. Held-out validation of the MLPs.** (a–c) DFT and MLP per-atom energies for maBN, maLiCl, and maBCN, respectively. (d–f) Corresponding Cartesian atomic-force components. The dashed lines indicate exact agreement.

*Figure S1 Alt Text:*

*Six DFT-versus-MLP parity plots for held-out maBN, maLiCl, and maBCN configurations. The upper row compares per-atom energies and the lower row compares Cartesian force components.*

*Most points lie close to the diagonal line of exact agreement, with energy errors of 8.3–12 millielectronvolts per atom and force errors of 0.046–0.21 electronvolts per angstrom.*

The configurations in Fig. S1 belong to the 10% validation subsets and were not used for optimization of the network weights. The per-atom energy RMSE values are 8.3, 8.5, and 12 meV atom$^{-1}$ for maBN, maLiCl, and maBCN, respectively; the corresponding force RMSE values are 0.13, 0.046, and 0.21 eV Å$^{-1}$. The close agreement of the parity distributions is the first validation level; the task-specific and finite-temperature tests below probe transfer to the structural-search and molecular-dynamics settings used in this work.

### B. DFT benchmarks of MC energy differences

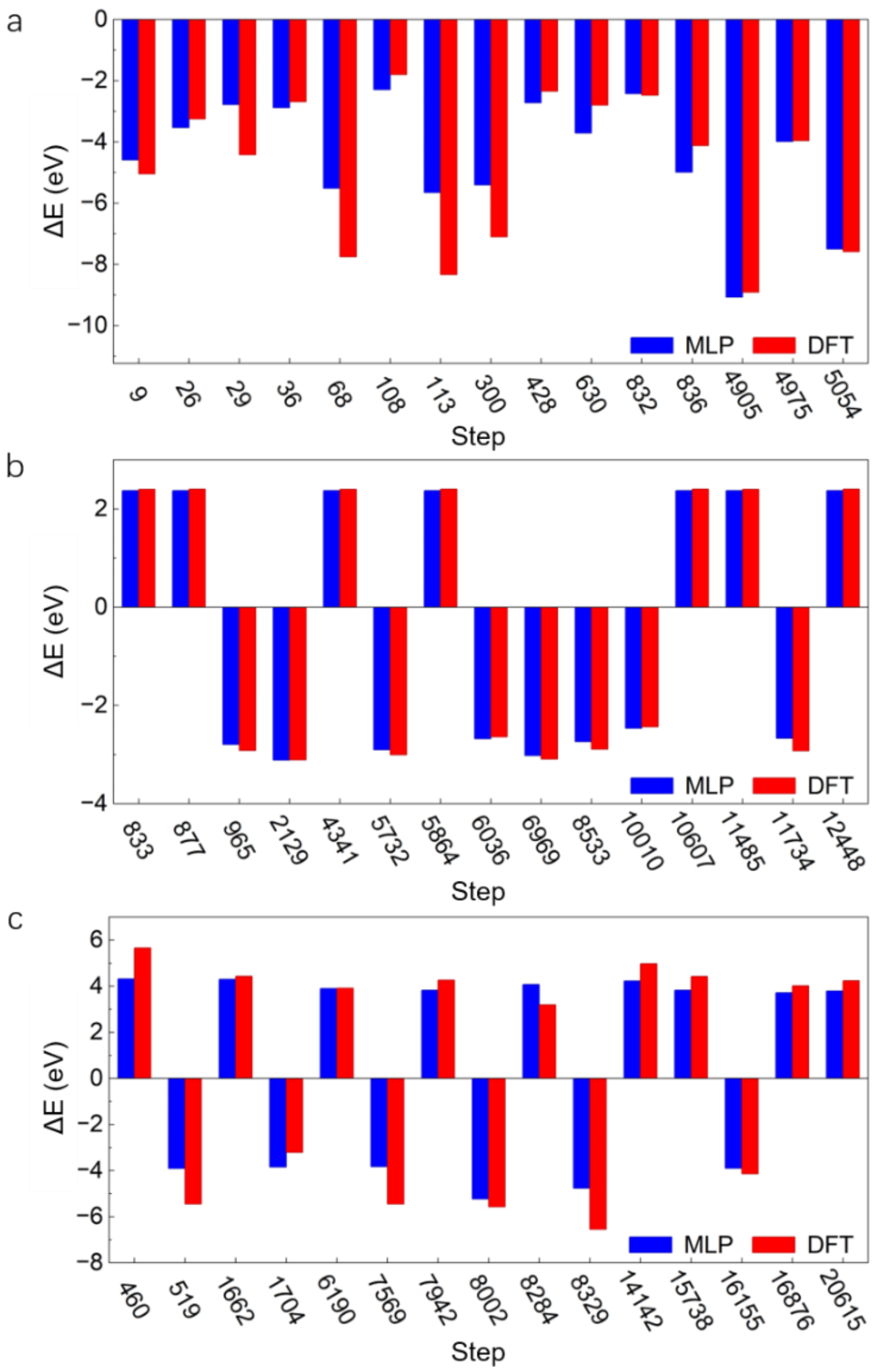

**Figure S2. Signed energy changes, $\Delta E = E_{after} - E_{before}$, predicted by the MLP and calculated by DFT for 15 representative EDKMC trial moves in each system: (a) maBN, (b) maLiCl, and (c) maBCN.** Positive and negative values denote energetically uphill and downhill moves, respectively; the horizontal line marks $\Delta E = 0$.

*Figure S2 Alt Text:*

*Three paired-bar charts compare the signed energy changes, $\Delta E = E_{after} - E_{before}$, predicted by the MLP and calculated by DFT for 15 representative EDKMC trial moves in each of maBN, maLiCl, and maBCN. Blue and red bars denote the MLP and DFT values, respectively, and the zero line separates energetically uphill moves ($\Delta E > 0$) from downhill moves ($\Delta E < 0$). All 15 sampled maBN moves are downhill; the maLiCl set contains 7 uphill and 8 downhill moves; and the maBCN set contains 9 uphill and 6 downhill moves according to DFT. The MLP and DFT assign the same sign to all 45 sampled moves, and the paired values are generally similar in magnitude.*

Because the MLP enters the structural search through the signed energy difference between the relaxed configurations before and after a trial move, $\Delta E = E_{after} - E_{before}$ is a task-specific validation quantity. Fifteen representative trial steps were selected for each 60-atom system, and the before/after configurations were recalculated using identical DFT settings. All 15 sampled maBN moves are downhill ($\Delta E < 0$); the maLiCl set contains 7 uphill and 8 downhill moves; and the maBCN set contains 9 uphill and 6 downhill moves according to DFT. The MLP and DFT assign the same sign to all 45 sampled moves and show close agreement in the magnitude of $\Delta E$ (Fig. S2). These results support the reliability of the MLP-predicted energetic direction for the sampled EDKMC trial moves.

### C. Finite-temperature DFT–MLP structural comparison

Independent NVT molecular-dynamics simulations were performed at 300 K with DFT and with the corresponding MLP, starting from the same 60-atom configuration for each material. Equivalent initial velocities, the same time step, and the same total duration were used. Each trajectory contained 8000 steps; the first 2000 steps were discarded and 100 snapshots were sampled at equal intervals from the remaining portion. These trajectory snapshots were not included in the training sets.

Partial in-plane radial distribution functions (RDFs), bond-length distributions, and bond-angle distributions were calculated from the same snapshot ensembles and with identical post-processing settings for DFT and MLP. The comparison therefore tests finite-temperature structural sampling rather than only pointwise energies and forces.

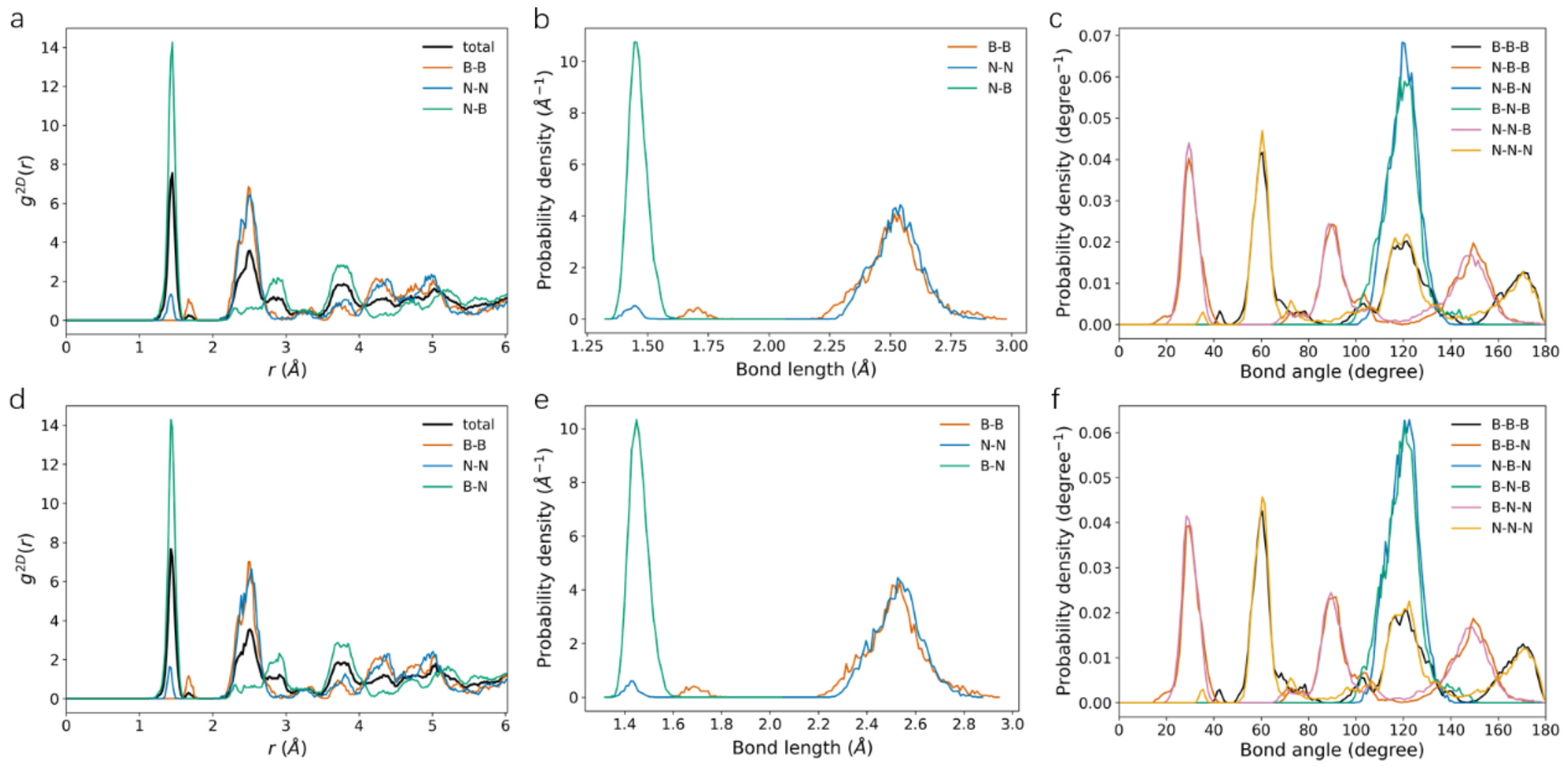

**Figure S3. Finite-temperature structural descriptors of the 60-atom maBN model at 300 K.** (a–c) Partial in-plane RDFs, bond-length distributions, and bond-angle distributions from DFT-MD; (d–f) the corresponding descriptors from MLP-MD.

*Figure S3 Alt Text:*

*Six plots comparing finite-temperature structural descriptors of a 60-atom maBN model at 300 kelvin. The upper row shows DFT molecular-dynamics partial radial distribution functions, bond-length distributions, and bond-angle distributions, while the lower row shows the corresponding MLP results. The major peak positions and overall distribution shapes are similar between the two methods.*

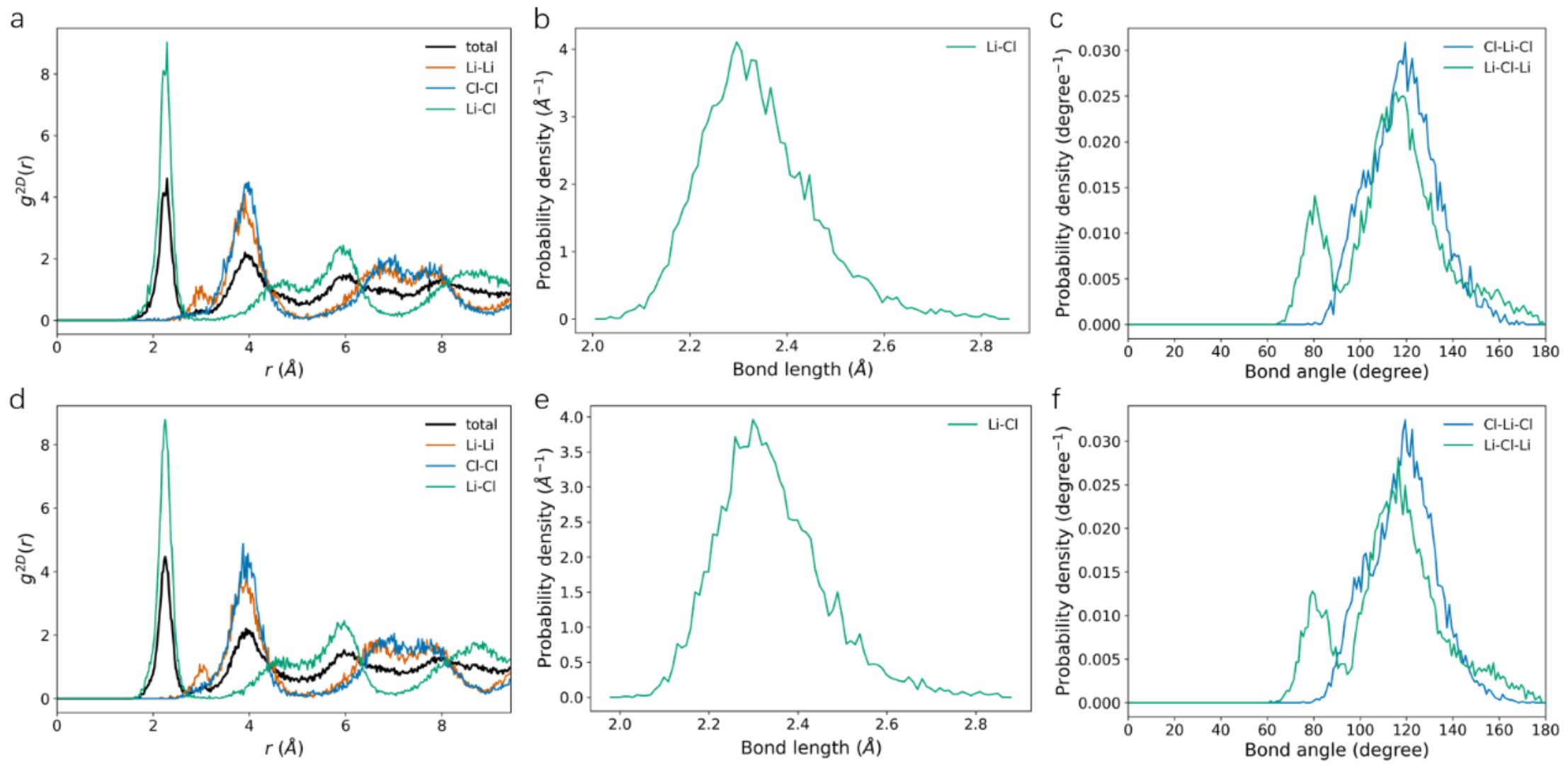

**Figure S4. Finite-temperature structural descriptors of the 60-atom maLiCl model at 300 K.** (a–c) Partial in-plane RDFs, bond-length distributions, and bond-angle distributions from DFT-MD; (d–f) the corresponding descriptors from MLP-MD.

*Figure S4 Alt Text:*

*Six plots comparing finite-temperature structural descriptors of a 60-atom maLiCl model at 300 kelvin. DFT molecular-dynamics results are shown in the upper row and MLP results in the lower row for partial radial distribution functions, Li–Cl bond lengths, and bond angles. The MLP reproduces the principal DFT peak positions and distribution shapes.*

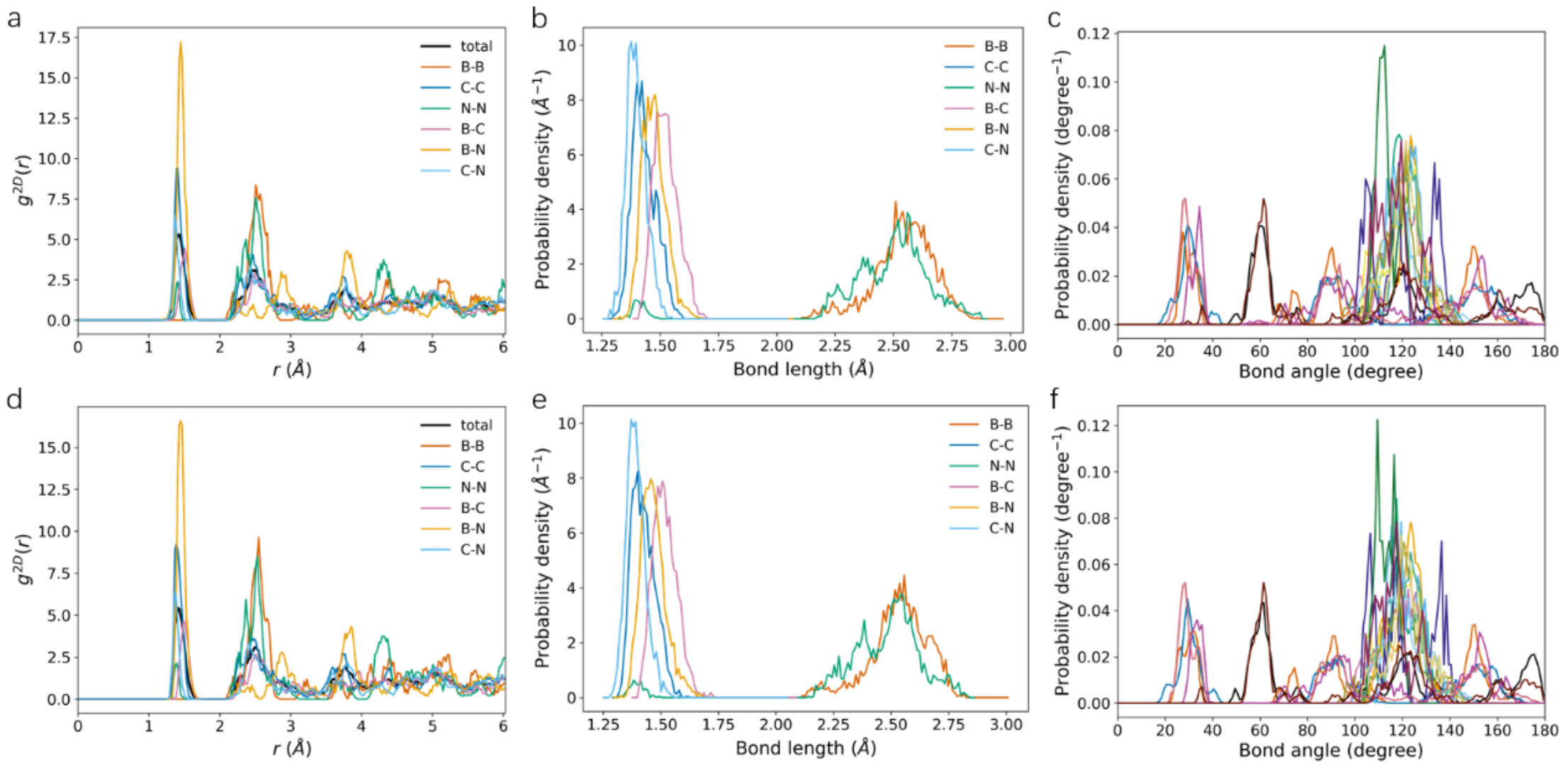

**Figure S5. Finite-temperature structural descriptors of the 60-atom maBCN model at 300 K.** (a–c) Partial in-plane RDFs, bond-length distributions, and bond-angle distributions from DFT-MD; (d–f) the corresponding descriptors from MLP-MD.

*Figure S5 Alt Text:*

*Six plots comparing finite-temperature structural descriptors of a 60-atom maBCN model at 300 kelvin. The upper and lower rows contain DFT and MLP results, respectively, for partial radial distribution functions, species-resolved bond-length distributions, and bond-angle distributions. Despite the many chemical environments, the dominant peaks and overall distributions are similar between the two methods.*

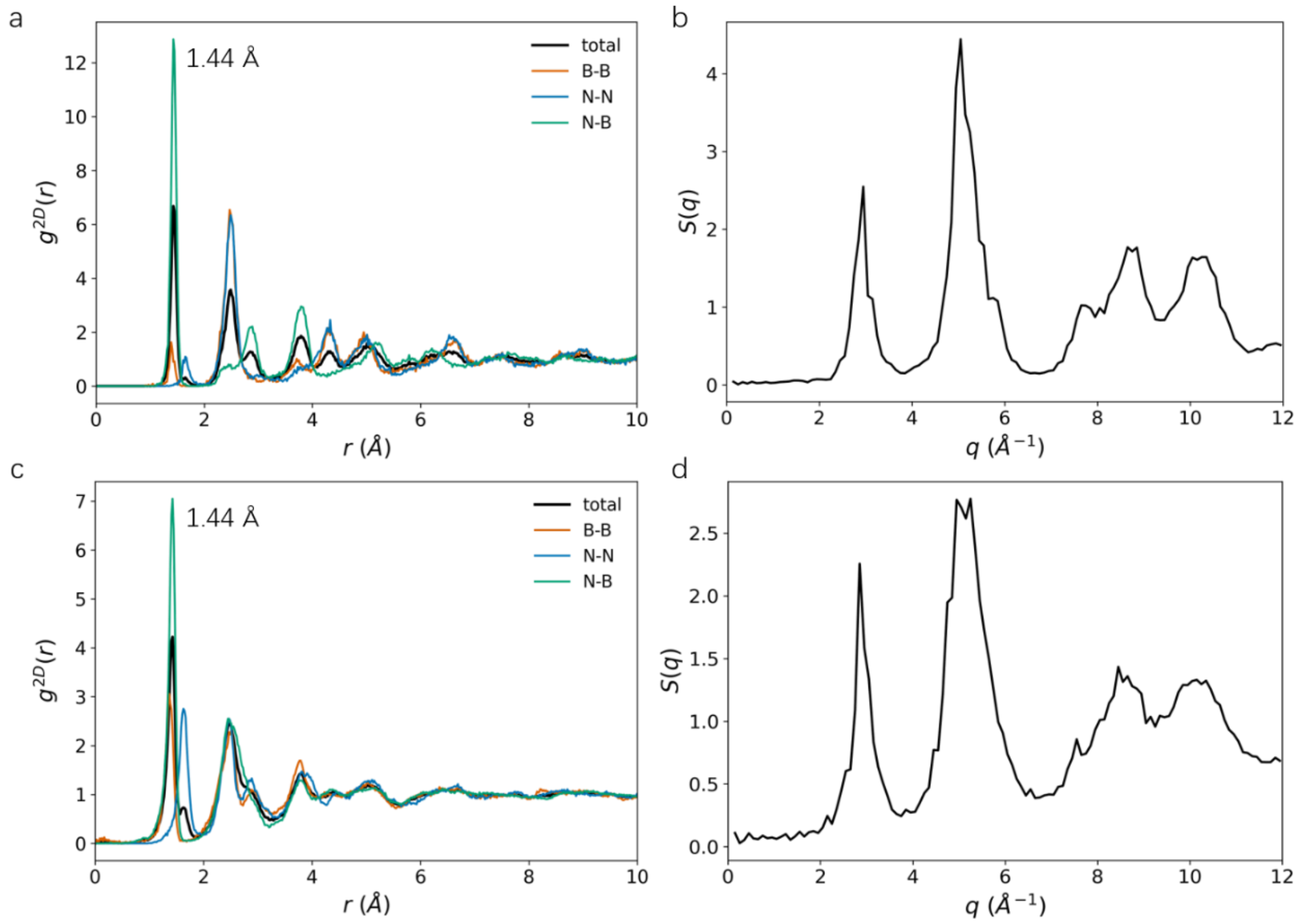

**Figure S6. Structural descriptors from 300 K MLP-MD simulations of large maBN models.** (a,b) Partial in-plane RDFs and S(q), respectively, for maBN generated by the MLP-driven structural search. (c,d) The corresponding quantities for maBN generated by the extended-Tersoff-driven structural search. All four panels were evaluated with the maBN MLP.

*Figure S6 Alt Text:*

*Four plots of structural descriptors calculated from 300-kelvin MLP molecular-dynamics trajectories of large maBN models. The upper row shows partial in-plane radial distribution functions and the structure factor for an MLP-search-generated structure, and the lower row shows the corresponding quantities for an extended-Tersoff-search-generated structure. The major peak positions and overall distribution shapes are similar between the two models. Both radial distribution functions have a first-neighbor peak near 1.44 angstroms.*

The DFT-MD and MLP-MD distributions in Figs. S3–S5 reproduce the same principal peak positions and overall line shapes for all three systems. For the larger maBN models, the dominant first-neighbor B–N distance is near 1.44 Å, consistent with the first peak reported for a finite-thickness amorphous BN film by Hong et al[8]. This comparison is qualitative because that experiment does not describe an isolated monolayer. Moreover, RDFs and S(q) are two-body descriptors and do not uniquely distinguish a continuous random network from mixed-MRO or crystallite-containing models.

## III. Structural definitions and domain-assignment protocol

A continuous random network (CRN) is defined here as a network without a spatially continuous domain that can be assigned to a crystalline reference tiling, although local coordination order may remain. A local motif is a chemically and topologically specified polygonal unit that can participate in a two-dimensional reference tiling. A crystallite is a connected domain formed by one motif class with a consistent local tiling relation. A pseudocrystallite-like region exhibits local or medium-range order but fails at least one of the complete reference-template, connectivity, or minimum-domain criteria. A polymorphic-crystallite structure contains two or more crystallographically distinct crystallite-domain types in the same amorphous network.

Local motifs are first identified from the nearest-neighbor bond graph used for the ring census. For maBN, six-membered rings are grouped under cyclic permutation, reversal, and B/N exchange: ABABAB is h-BN-like; ABBABB/AABAAB is o-$B_2N_2$-like; and ABBAAB/AABABB is o-$B_4N_4$-like. The maLiCl library contains chemically alternating hexagons and squares, and the maBCN library contains graphene-like C hexagons, h-BN-like hexagons, borophene-like B triangles, and the additional chemically decorated motifs shown in Fig. S7. Local motif identity alone is not interpreted as a crystallite domain.

A polygon adjacency graph is then constructed by joining two motifs only when they share a complete edge. The minimum accepted patches are those illustrated in Fig. S7: three $B_3$ triangles connected through two shared edges; three LiCl squares forming an L-shaped patch with two shared edges; or three compatible hexagons forming the compact three-hexagon patch in which the motifs are pairwise edge-sharing. Isolated motifs and two-motif clusters are excluded. Larger domains are connected components that contain an accepted minimum patch and extend it through edge-sharing motifs consistent with the same reference tiling. Components containing more than one accepted motif class are assigned to mixed medium-range order (MRO) rather than to a phase-pure crystallite.

For each accepted component, the projected polygon area is computed from its in-plane atomic vertices. Shared edges are counted once by taking the union of all polygon areas. Let $A_{MRO}$ be the union area of all accepted phase-pure and mixed-MRO components and $A_{tot}$ the projected simulation-cell area. The connected structural-order fraction is $F_{MRO} = \frac{A_{MRO}}{A_{tot}}$ and the structural degree of disorder is $DOD_{str} = 1 - F_{MRO}$. Phase-resolved fractions are reported separately; for maBN, $F_{MRO} = F_{h-BN} + F_{o-B_2N_2} + F_{o-B_4N_4} + F_{mix}$. The same procedure is applied using the appropriate motif library for maLiCl and maBCN. Unlike the DOD employed in our earlier work[9], $DOD_{str}$ is used here only as a geometrical structural descriptor; no direct relation to thermal, mechanical, or other physical properties is assumed.

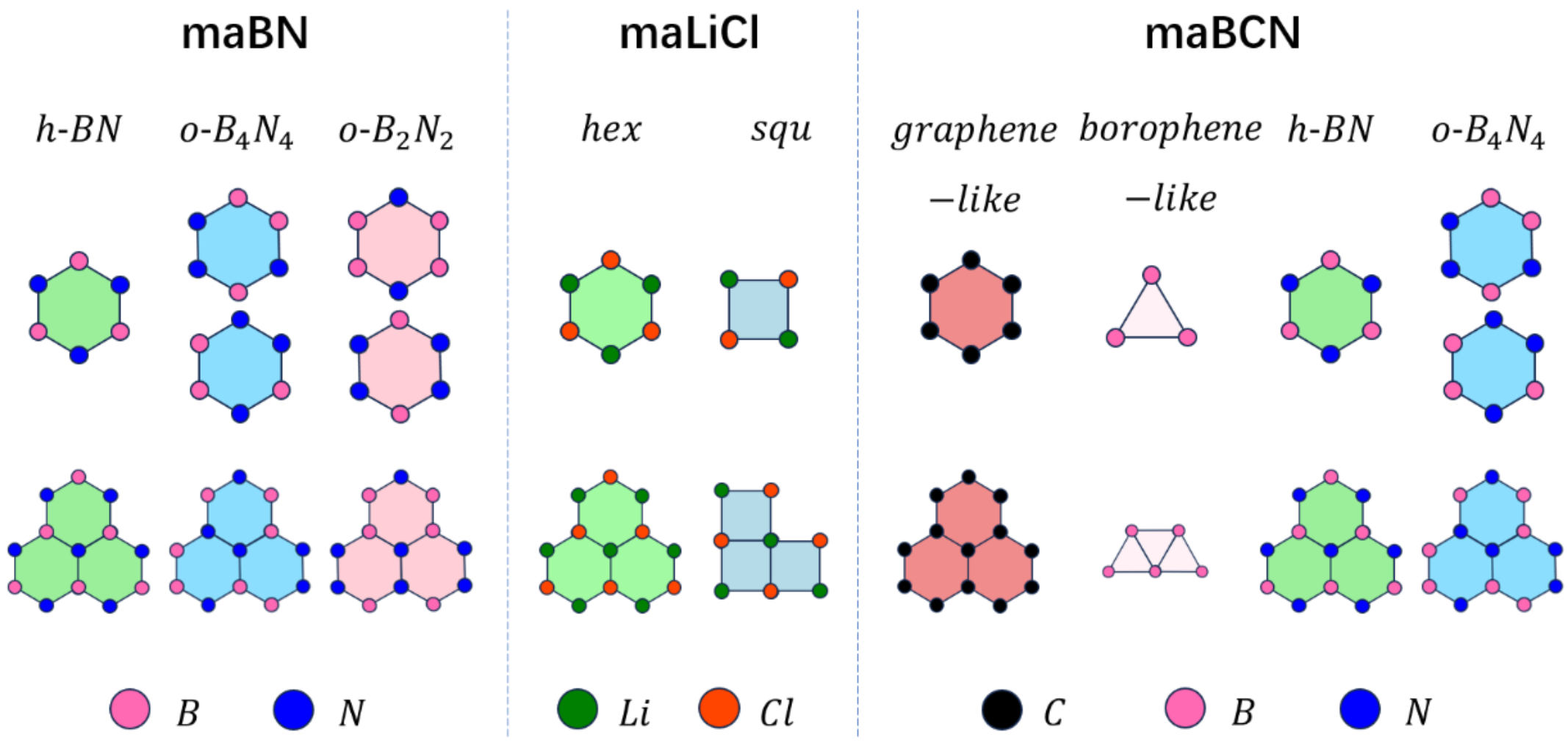

**Figure S7. Local-motif and minimum-domain definitions for maBN (left), maLiCl (middle), and maBCN (right).** The upper rows show the motif libraries; the lower rows show the smallest accepted edge-connected patches. Atom colors follow the manuscript convention. Isolated motifs and two-motif clusters are not counted as MRO domains.

*Figure S7 Alt Text:*

*Schematic libraries of local motifs and minimum accepted connected domains for maBN, maLiCl, and maBCN. The upper row shows the individual reference motifs, including h-BN-like and noncanonical BN rings, hexagonal and square LiCl motifs, and graphene-like, borophene-like, h-BN-like, and noncanonical BCN motifs. The lower row shows the smallest three-motif edge-connected patches counted as medium-range-order domains; isolated motifs and two-motif clusters are excluded.*

## IV. Independent DFT validation of the two maBN ensembles

To avoid evaluating MLP-generated and extended-Tersoff-generated structures only on the MLP energy surface, we selected 15 equilibrated configurations from each search at the same MC acceptance parameter. The 30 configurations were evaluated by DFT single-point calculations using identical settings and without further structural relaxation.

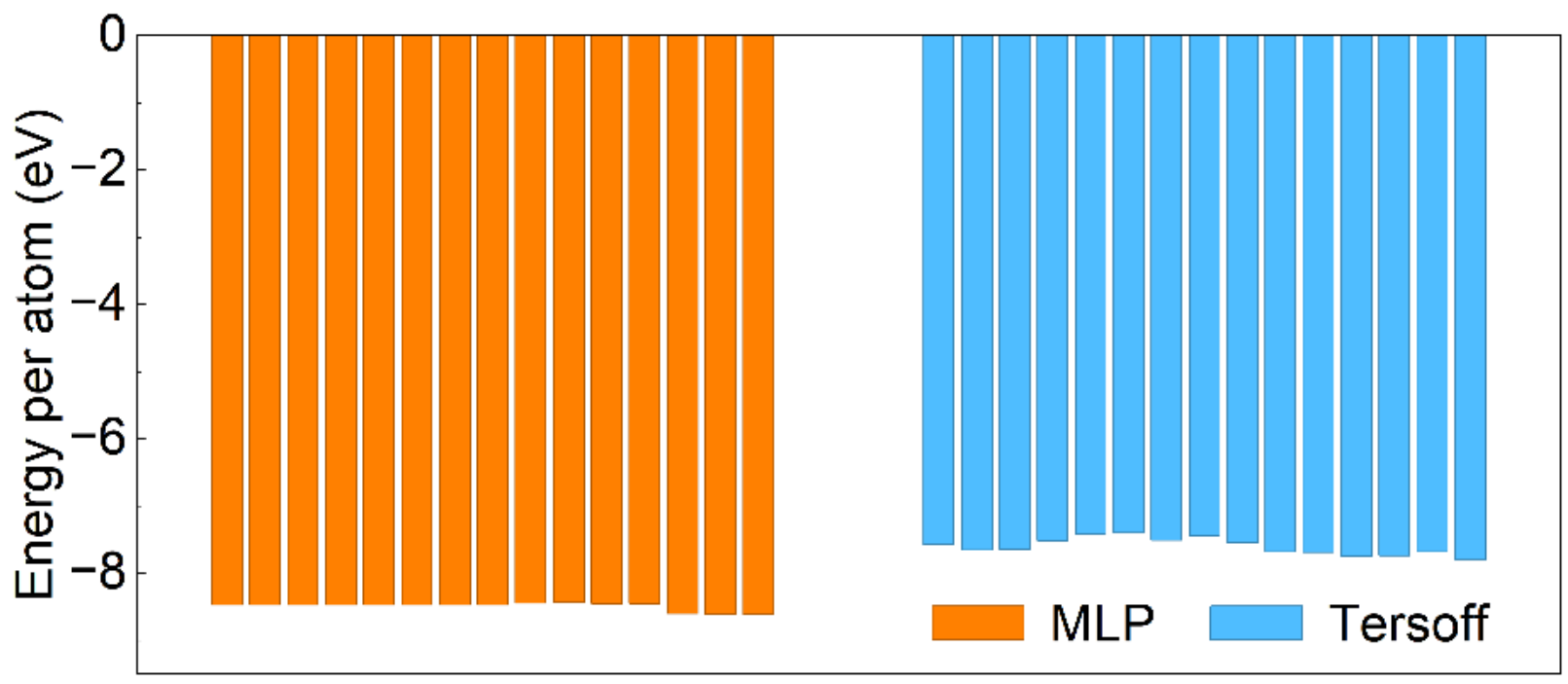

**Figure S8. Independent DFT single-point energies of 15 maBN structures generated by the MLP-driven search and 15 generated by the extended-Tersoff-driven search.** Bars show individual configurations selected at the same MC acceptance parameter.

*Figure S8 Alt Text:*

*Bar chart of DFT single-point energies for 15 MLP-search-generated and 15 extended-Tersoff-search-generated maBN configurations. The MLP-generated configurations form a lower-energy group, with a mean of minus 8.49 electronvolts per atom, whereas the extended-Tersoff-generated group has a mean of minus 7.60 electronvolts per atom.*

The MLP-generated ensemble has a DFT mean energy of −8.49 ± 0.06 eV atom$^{-1}$, whereas the extended-Tersoff-generated ensemble has −7.60 ± 0.13 eV atom$^{-1}$ (mean ± standard deviation). The lower mean energy of the MLP-generated structures is therefore confirmed independently at the DFT level and is not an artifact of scoring both ensembles with the MLP.

**Table S1. Energies of the periodic o-$B_2N_2$ and o-$B_4N_4$ reference phases relative to h-BN, calculated using DFT, the maBN MLP, and the extended Tersoff potential.**

*Table S1 Alt Text:*

*Relative energies of periodic o-$B_2N_2$ and o-$B_4N_4$ reference phases with respect to h-BN. DFT and the MLP place both noncanonical phases above h-BN and give similar values: 1.65 versus 1.63 electronvolts per atom for o-$B_2N_2$ and 0.81 versus 0.84 electronvolts per atom for o-$B_4N_4$. The extended Tersoff potential instead places both phases below h-BN, reversing the DFT energetic ordering.*

| Energy (eV/atom) | DFT | MLP | Tersoff |
|---|---|---|---|
| E(o-$B_2N_2$)-E(h-BN) | 1.65 | 1.63 | -0.51 |
| E(o-$B_4N_4$)-E(h-BN) | 0.81 | 0.84 | -0.27 |

The MLP reproduces the DFT energetic ordering and relative energy differences of the three periodic reference phases. In contrast, the extended Tersoff potential reverses the ordering and places the noncanonical o-$B_2N_2$ and o-$B_4N_4$ phases below h-BN. This behavior indicates an over-stabilization of the B–B/N–N-containing environments by the empirical potential. The periodic-phase comparison explains the distinct motifs favored by the two searches, while the ensemble benchmark in Fig. S8 provides the direct amorphous-structure validation.

## V. Effect of the initial structure on the MC-search outcome

Because a finite structural search can retain memory of its starting topology, we repeated the calculations from alternative initial configurations. These controls test whether the stationary energy range is inherited solely from the crystalline or compositionally separated starting structures; they do not establish a unique equilibrium amorphous state.

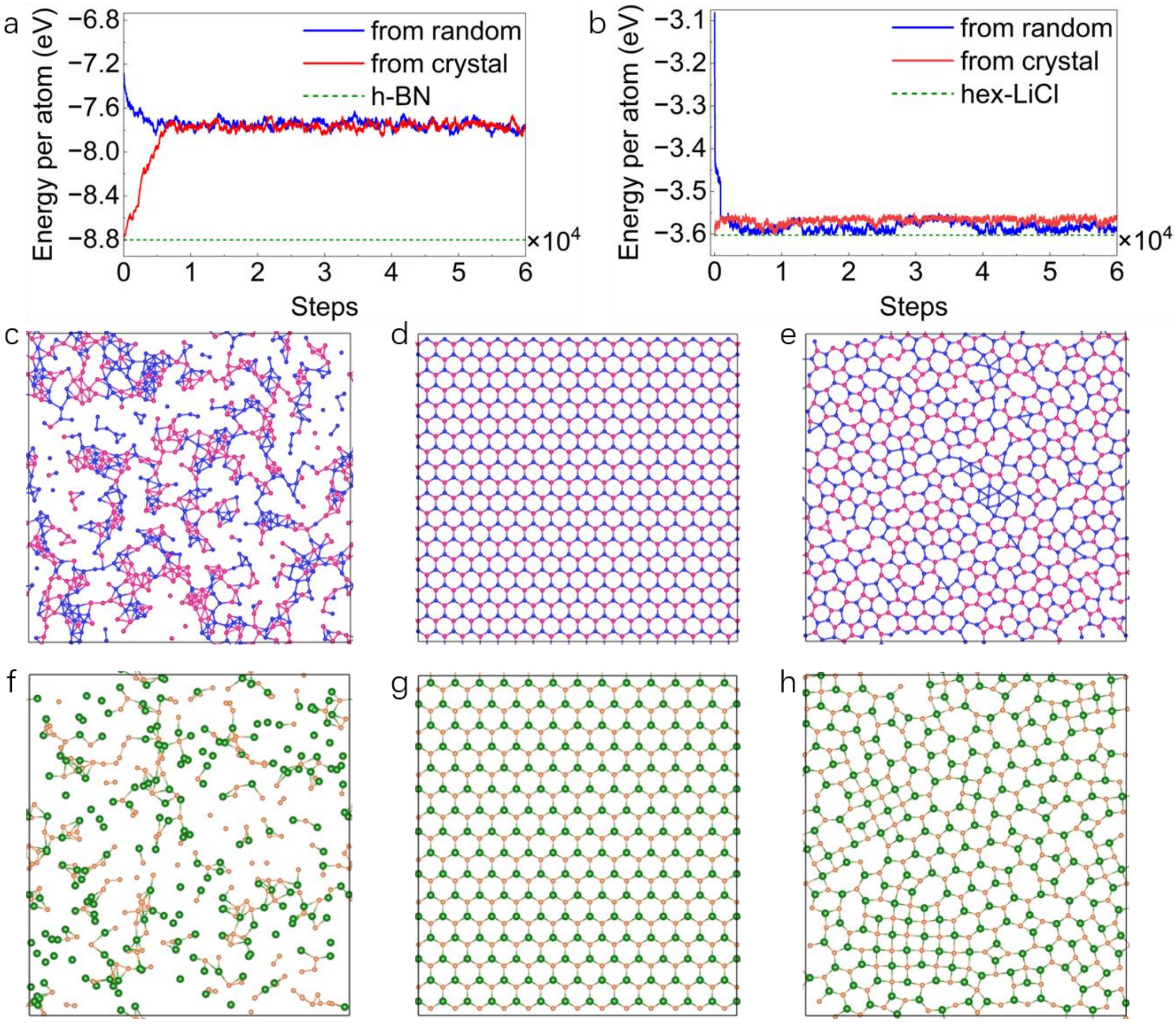

**Figure S9. EDKMC structural searches for maBN and maLiCl from different initial structures.** (a,b) Energy per atom versus MC step for maBN and maLiCl, respectively, starting from random and crystalline configurations; dashed lines mark the h-BN and hexagonal-LiCl reference energies. (c–e) Random initial, h-BN initial, and representative final maBN structures. (f–h) Random initial, hexagonal-LiCl initial, and representative final maLiCl structures.

*Figure S9 Alt Text:*

*Energy trajectories and representative structures from maBN and maLiCl EDKMC searches initiated from random and crystalline configurations. For both materials, searches from the two*

*starting structures approach similar stationary energy ranges. The accompanying atomic configurations show the random and crystalline initial states and representative disordered final structures.*

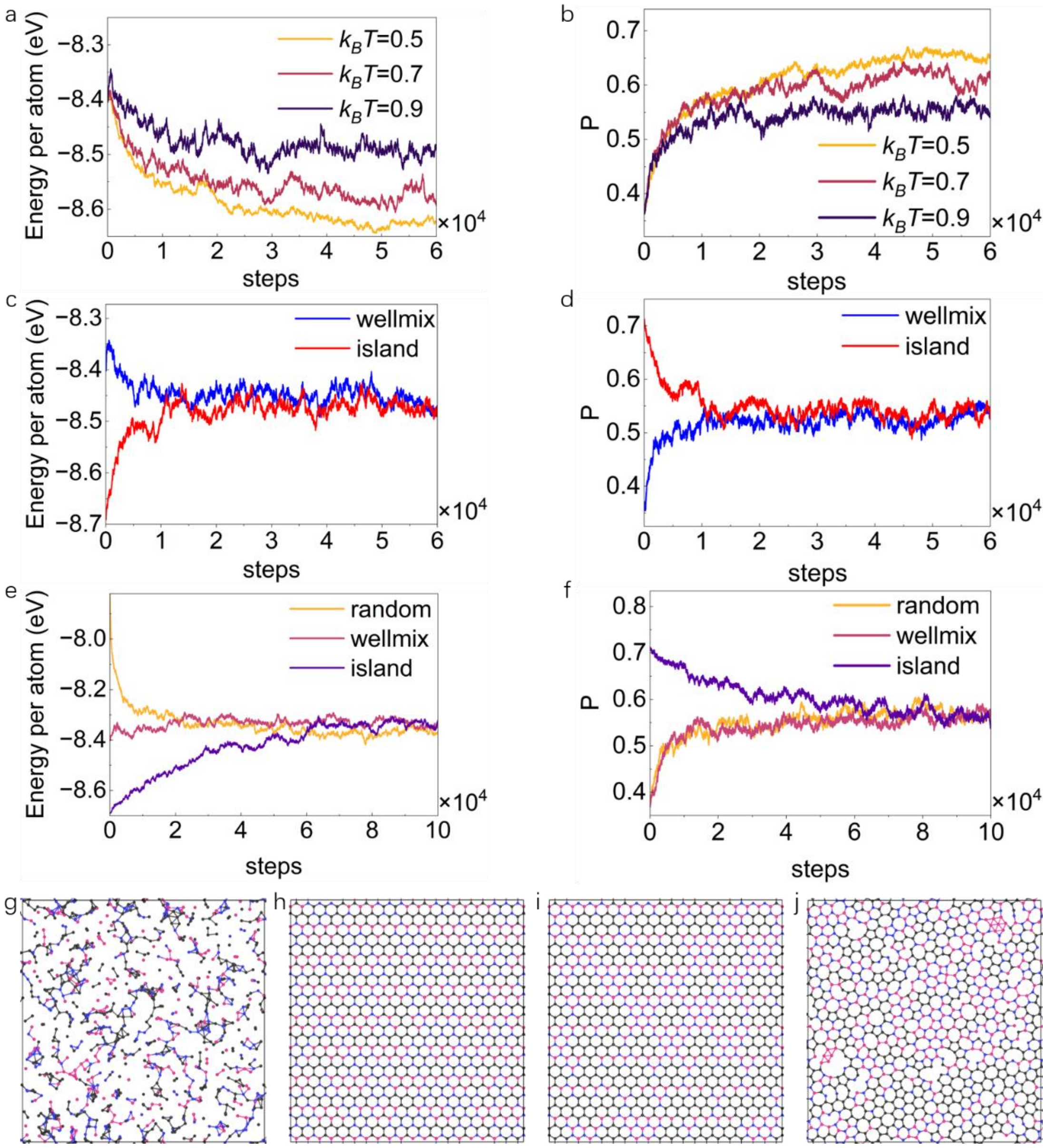

**Figure S10. Dependence of the maBCN structural search on the acceptance parameter and initial chemical arrangement.** (a,b) Energy per atom and chemical-order parameter P versus MC step for exchange-only searches from the same well-mixed structure at acceptance parameters 0.5, 0.7, and 0.9 eV. (c,d) Exchange-only searches at 1.0 eV from well-mixed and island-like starts. (e,f) Full searches using 50% atom-exchange and 50% Stone–Wales moves at 0.7 eV from random, well-mixed, and island-like starts. (g–j) Random, well-mixed, and island-like initial structures and a representative final maBCN structure, respectively. The legend symbol $k_BT$ denotes the algorithmic acceptance-energy parameter and not a thermodynamic temperature.

*Figure S10 Alt Text:*

Ten-panel comparison of maBCN structural searches using different acceptance-energy parameters, initial chemical arrangements, and trial-move sets. Energy and chemical-order trajectories show parameter-dependent behavior in exchange-only searches, while well-mixed and island-like starts approach similar ranges at an acceptance parameter of 1.0 electronvolt. In full searches combining atom exchanges and Stone–Wales transformations, random, well-mixed, and island-like initial structures progressively approach similar energy and chemical-order ranges; the bottom row displays the three initial structures and a representative final amorphous configuration.

For maBCN, $P = \frac{N_{C-C} + N_{B-N}}{N_{total}}$ is used as a chemical-order descriptor, where $N_{C-C}$ and $N_{B-N}$ are the numbers of C–C and B–N bonds and $N_{total}$ is the number of atoms. A larger P corresponds to a greater population of C–C and B–N connections. In the exchange-only tests, energy and P approached stationary ranges after approximately $5 \times 10^4$ steps. At fixed acceptance parameter, the well-mixed and island-like starts converged to similar energy and P ranges, whereas increasing the acceptance parameter produced a higher-energy, more mixed stationary ensemble.

In the full searches at an acceptance parameter of 0.7 eV, random, well-mixed, and island-like starts began at distinct energies but approached the same stationary energy range. The well-mixed and island-like trajectories also reached similar P values after approximately $7 \times 10^4$ steps. The maBN and maLiCl tests in Fig. S9 show analogous convergence of the energy trajectories from random and crystalline starts. These results reduce the likelihood that the reported energetic tendencies are imposed solely by a single initial structure under the tested settings, while the detailed finite-search pathways and domain morphologies remain history dependent.